\documentclass{aa}
\usepackage{graphicx}
\usepackage{epsfig}
\usepackage{natbib}
\bibpunct{(}{)}{;}{a}{}{,}

\def\ms{\,m\,s$^{-1}$}   

\begin{document}


\title{The CORALIE survey for southern extra-solar planets VII.}
\subtitle{Two short-period Saturnian companions to \object{HD\,108147} and
  \object{HD\,168746}\thanks{Based on observations collected with the
    {\footnotesize CORALIE} echelle spectrograph on the 1.2-m Euler
    Swiss telescope at La\,Silla Observatory, ESO Chile}}

\author{F.~Pepe \and M.~Mayor \and F.~Galland \and D.~Naef \and D.~Queloz \and N.C. Santos \and S.~Udry \and M. Burnet }

\offprints{F. Pepe, \email{Francesco.Pepe@obs.unige.ch}}

\institute{Observatoire de Gen\`eve, 51 ch. des Maillettes, CH--1290 Sauverny, Switzerland}

\date{Received / Accepted } 

\abstract{ We present the discovery of two Saturn-mass companions to \object{{\footnotesize HD}\,108147} and \object{{\footnotesize HD}\,168746}. Both belong to the lightest ever discovered planets. The minimum mass of the companion to \object{{\footnotesize HD}\,168746} is of only 0.77 the mass of Saturn and its orbital period is 6.4 days. The companion to \object{{\footnotesize HD}\,108147} orbits its parent star in 10.9 days and its minimum mass is 1.34 that of Saturn. Its orbit is characterized by a high eccentricity, $e=0.50$, indicating possibly the presence of a second companion. \par
The detection of Saturn-mass planets by means of the Doppler technique demands high radial-velocity measurement precision. The two new candidates were discovered by means of the {\footnotesize CORALIE} echelle spectrograph. The instrumental accuracy of {\footnotesize CORALIE} combined with the simultaneous ThAr-reference technique has reached a level better than 3\,\ms. On many observed objects the precision is now limited by photon noise. We present in this paper the weighted cross-correlation technique, which leads to an improvement in the photon noise of the computed radial velocity. We discuss as well a modification of the numerical cross-correlation mask which reduces significantly the residual perturbation effects produced by telluric absorption lines.

}

\maketitle

\section{Introduction}
To date about 80 planetary candidates have been discovered. Their orbits show a widely-spread distribution of orbital parameters such as eccentricity and orbital period \citep{Udry-2001} which might be related to different evolution mechanisms. The minimum mass of the detected planetary companions ranges from $\sim$\,10 Jupiter masses down to about half the mass of Saturn. Four of the known candidates have sub-saturnian masses \citep{Jorissen-2001}. The lightest planet, \object{{\footnotesize HD}\,83443\,c}, has been discovered by means of the {\footnotesize CORALIE} spectrograph and has a mass of only 0.53\,M$_{\rm Sat}$ \citep{Mayor-2000:b}.
\par
The technique which underlies all these discoveries is that of precise radial-velocity measurements. It provides us with most orbital parameters. Because this technique does not allow us to constrain the angle of projection $\sin i$ of the orbital plane, only a minimum mass $m_2\sin i$ of the companion can be determined. The radial-velocity variation induced by a planetary companion on its parent star is large if its minimum mass $m_2\sin i$ is large. This fact makes the technique particularly sensitive to high-mass companions. On the other hand, the detection of low-mass planets is more difficult and produces a detection bias on the low-mass end of the mass function \citep[e.g.][]{Udry-2001,Jorissen-2001}. In order to reduce this bias the measurement precision must be improved by efficient and stable instrumentation and by efficient measurement and data reduction techniques.
\par
Using the {\footnotesize CORALIE} echelle spectrograph on the \mbox{1.2-m} Euler Swiss telescope at La\,Silla we are carrying out, since summer 1998, a large high-precision radial-velocity program \citep{Queloz-2000:b,Udry-2000:a}. Together with {\footnotesize ELODIE} in the northern hemisphere, {\footnotesize CORALIE} has allowed the discovery of about half of the known exoplanet candidates \citep{Udry-2001}. Recent improvement of the data reduction software have contributed to obtain an overall \emph{instrumental} precision below $\sim$\,3\,\ms\ \citep{Queloz-2001:a} over time scales longer than 3~years. The recent asteroseismology measurements on $\alpha$\,Cen\,A \citep{Bouchy-2001} have proven that the short term precision is even better, namely about 1\,\ms\ r.m.s over 1~night. The limitations are now determined mainly by photon noise and residual astroclimatic influence.
\par
The present paper describes the discovery of two Saturn-mass companions to the stars \object{{\footnotesize HD}\,108147} and \object{{\footnotesize HD}\,168746}. \object{{\footnotesize HD}\,168746}\,b, with its minimum mass of only 0.77\,M$_{\rm Sat}$, is one of the four sub-saturnian planets discovered to date. On the other hand, \object{{\footnotesize HD}\,108147}\,b distinguishes itself by a high eccentricity, fueling the discussion of the origin of such high eccentricity in the case of short-period companions. Before the orbital parameters and the stellar characteristics of these two objects described in detail in the second part of the paper, we discuss additional improvements made recently in the data reduction, and more precisely, in the way of extracting the radial-velocity information obtained from the high-resolution spectra recorded with {\footnotesize CORALIE}. 
  
\section{Improving the radial-velocity precision on {\footnotesize CORALIE} data}
\subsection{The weighted cross-correlation}
The technique of the numerical cross correlation used to determine the precise radial velocity is described in detail in \citet{Baranne-96} and will not be discussed here. We only recall that the measured spectrum is correlated with a numerical mask consisting of 1 and 0 value-zones, with the non-zero zones corresponding to the theoretical positions and widths of the stellar absorption lines at zero velocity. The cross-correlation function (CCF) is constructed by shifting the mask as a function of the Doppler velocity:
\begin{eqnarray}
CCF(v_{\rm R}) & = & \int S(\lambda) \cdot M(\lambda_{v_{\rm R}})\,d\lambda\\
& = & \int S(\lambda) \cdot \sum_i M_i(\lambda_{v_{\rm R}})\,d\lambda\\
& = & \sum_i \int S(\lambda) M_i(\lambda_{v_{\rm R}})\,d\lambda = \sum_i CCF_i(v_{\rm R}),\\
{\rm where} & & \nonumber\\
\lambda_{v_{\rm R}} & = & \lambda \sqrt{\frac{1-\frac{v_{\rm R}}{c}}{1+\frac{v_{\rm R}}{c}}}~.
\end{eqnarray}
In this equation $S(\lambda)$ is the recorded spectrum, and $M(\lambda_{v_{\rm R}})$ represents the Doppler-shifted numerical mask which can be expressed as the sum of masks $M_i$ each corresponding to a stellar absorption line $i$. The resulting $CCF(v_{\rm R})$ is a function describing somehow a flux-weighted ``mean'' profile of the stellar absorption lines transmitted by the mask. For the radial velocity value of the star, we take the minimum of the CCF, fitted with a Gaussian function.
\par
The cross-correlation technique has proven to be very robust and simple, and to deliver excellent results. The great advantage of the cross-correlation with a numerical mask is that it does not need any high signal-to-noise reference spectra to compute the precise radial velocity. However, it has been shown in the past \citep{Bouchy-2001b,Chelli-2000} that in terms of photon noise the technique could still be improved. In fact, the CCF does not extract the radial-velocity information content in an optimized way. Deep and sharp lines, for example, are not weighted sufficiently, although they contain more radial-velocity information than broad and weak lines. The challenge was therefore to combine the simplicity and robustness of the cross correlation with a more efficient information extraction.
\par
The solution presented here assumes that the absorption lines of the stellar spectrum used to compute the radial velocity have all about the same FWHM, which is a good approximation since broad lines are not included in our standard numerical mask. The lines have different relative depths which are averaged in the resulting CCF. Lines with large relative depth contain intrisically more radial-velocity information than weak lines, however. Our goal is therefore to build up a weighted cross-correlation function $CCF^{\rm w}$ which accounts for the correct weight of each spectral line contained in the mask:
\begin{eqnarray}
\label{eq:weight1}
CCF^{\rm w}(v_{\rm R}) & = & \sum_i CCF_i(v_{\rm R}) \cdot w_i
\end{eqnarray}
In order to understand the origin of $w_i$ the following example shall be presented: The contribution  to the global CCF of a spectral line $l$ with relative depth $c_l=0.5$ and a continuum level of $S_l=I_0$ is exactly the same as for a line $m$ with relative depth $c_m=1$ but a continuum level at $S_m=I_0/2$. The cross-correlation signal $CCF_i \propto c_i \cdot S_i$ is indeed of equal amplitude in both cases, namely $CCF_l = CCF_m \propto I_0/2$. The noise on each point of the cross-correlation function is however proportional to the square root of the real spectral signal $S_i$, and thus $\sqrt{2}$ times larger in the $CCF_l$. If we would fit each cross-correlation function $CCF_i$ independently we would get an precision on the fitted position which is $\sqrt{2}$ larger for the line~$m$. Thus, the weight of the line $l$ should be greater.
\par
In general, for a given amplitude of the cross-correlation signal $CCF_i$, the noise on each point of the CCF -- and thus on the resulting Gaussian fit -- is $\sigma_i^2 \propto S_i \propto \frac{CCF_i}{c_i}\propto \frac{1}{c_i}$. The weight we have to give to the single $CCF_i$ must therefore be equal to $w_i=\frac{1}{\sigma_i^2}=c_i$, where $c_i$ is the relative depth of each spectral line $i$. Eq.~(\ref{eq:weight1}) becomes then:
\begin{eqnarray}
\label{eq:weight2}
CCF^{\rm w}(v_{\rm R}) & = & \sum_i CCF_i(v_{\rm R}) \cdot c_i\\
\label{eq:weight3}
& = & \int I(\lambda) \cdot \sum_i (M_i(\lambda_{v_{\rm R}}) \cdot c_i) \,d\lambda~.
\end{eqnarray}
By means of a synthetic K0 dwarf spectrum we have computed the relative depth $c_i$ (and thus the weight) for each absorption line contained in the numerical mask. We have modified the reduction software to take into account the relative weight of each line when computing the CCF. Finally, we have compared the old and the new code by numerical simulation on a set of 100 spectra issued from a high signal-to-noise spectrum on which statistical noise has been added. The result was a reduction of 1.25 of the measured radial velocity dispersion which is equivalent to an hypothetical increase in signal-to-noise on the spectrum of the same value. We have also applied the old and the new algorithm on the radial-velocity data of {\footnotesize HD\,108147} and {\footnotesize HD\,168756}. The results of this comparison will be presented below.

\subsection{Telluric lines: reduced contamination effects}
\begin{figure}
\includegraphics[width=\hsize]{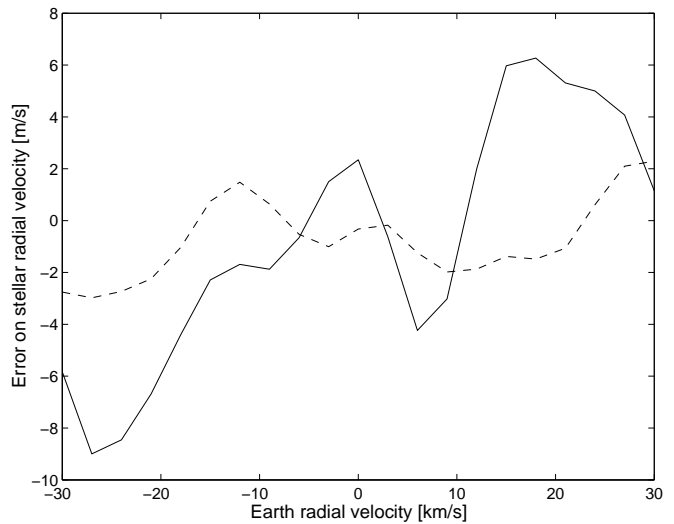}
\caption{The plot shows the error on the measured radial velocity produced by telluric lines as a function of Earth radial velocity. For this purpose a real stellar spectrum has been Doppler-shifted by a radial-velocity value ranging from $-30$\,km\,s$^{-1}$ to $+30$\,km\,s$^{-1}$ and superimposed with a zero-velocity telluric absorption spectrum. The obtained radial velocity has then been compared to the star velocity obtained on the original stellar spectrum, the error being the difference between them. The solid line shows the results using the original numerical mask while the dashed line represents the results with the new clean mask}
\label{fi:atm_error}
\end{figure}

Telluric absorption lines superimposed to the stellar spectrum form absorption bands, especially in the red-visible wavelength region. The position of the telluric lines relative to the stellar lines changes as a function of the relative radial velocity between Earth and star, while the strength of the telluric lines depends strongly on the astroclimatic conditions. When cross-correlating the recorded spectrum with the numerical mask a parasitic signal is produced by the \emph{casual} correlation of a line in the mask with a telluric line in the spectrum. This might introduce an asymmetry in the {\footnotesize CCF} and consequently an error on the measured stellar radial velocity. In principle, the cross-correlation mask used in the {\footnotesize ELODIE} and {\footnotesize CORALIE} data reduction has been ``cleaned'' from telluric lines: all stellar lines for which the wavelength lies close (at a couple of line widths) to that of a telluric line have been removed from the numerical mask. Because of the quite large annual Earth motion, however, the telluric lines change their relative position with regard to the stellar lines by about $\pm 30$\,km\,s$^{-1}$ ($\pm 0.6$\,\AA~at~6000\,\AA). These means that every stellar line lying within $\pm 30$\,km\,s$^{-1}$ from a telluric line is potentially affected by a telluric line more or less strongly in some period of the year. The superposition of these effects can then result in an increased ``noise'' or even in an erroneous periodic signal on the stellar velocity with a period of one year. Fig.~\ref{fi:atm_error} shows simulations made on real {\footnotesize ELODIE} spectra.
\par

In order to reduce this effect we have cleaned the original mask carefully from \emph{all} stellar lines potentially affected by telluric lines. Although the photon-noise was slightly increased -- since some regions of the stellar spectra were not used anymore in the cross correlation -- quite impressive results have been obtained. In our simulations and as shown by the dashed curved in Fig.~\ref{fi:atm_error}, the error produced by the telluric lines has been reduced at least by a factor of 3.
\par
Although in case of the {\footnotesize CORALIE} instrument the astroclimatic conditions are in average more convenient than on {\footnotesize ELODIE}, and therefore the effect of the telluric lines is expected to be less important, we have decided to apply the new numerical mask on the radial velocity data of {\footnotesize HD\,108147} and {\footnotesize HD\,168756} presented in this paper. The corresponding results are discussed below.
\par Finally, we should also mention that the improvement of the mask presented in this section does not apply only to newly acquired data. It will allow us to re-correlate \emph{any} spectrum obtained with {\footnotesize ELODIE} or {\footnotesize CORALIE} until the present and improve the existing radial-velocity data base.

\section{A Saturn-mass planet in eccentric orbit around \object{{\footnotesize HD}\,108147}}

\subsection{Stellar characteristics of \object{{\footnotesize HD}\,108147}}

\object{{\footnotesize HD}\,108147} ({\footnotesize HIP}\,60644) is a F8/G0 dwarf in the Crux constellation. Its magnitude is $V$\,=\,6.99 while the {\footnotesize HIPPARCOS} catalogue \citep{ESA-97} lists a color index $B-V$\,=\,0.537. The precise astrometric parallax is $\pi=25.95\pm 0.69$\,mas corresponding to a distance of about 38.57\,pc from the Sun. The derived absolute magnitude, $M_V$\,=\,4.06, is typical for a G0 dwarf.
\par
Stellar parameters such as effective temperature $T_{\rm eff}=6265$\,K, surface gravity $\log g=4.59$, as well as [Fe/H]\,=\,$+0.2$ have been derived in the detailed LTE spectroscopic analysis carried out by \citet{Santos-2001:a}. The obtained metallicity is slightly higher than the average value for stars of the {\footnotesize CORALIE} sample, like most of the stars with giant planets, and is very close to the mean value of [Fe/H] of stars with planets \citep{Santos-2001:a}. Using the evolutionary tracks of the Geneva models given by \citet{Schaerer-93}, \citet{Santos-2001:a} compute a stellar mass $M=1.27$\,M$_{\odot}$. The stellar mass is higher than the ``typical'' mass of G0 dwarfs and can be explained by the high metallicity of the star.

\begin{figure}
\includegraphics[width=\hsize]{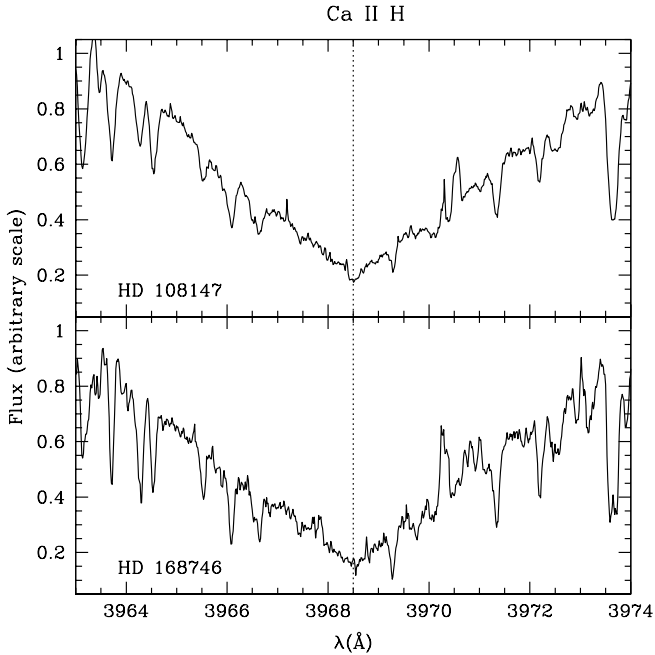}
\caption{$\lambda 3968.5$\,\AA~\ion{Ca}{ii}\,H absorption line region of the summed {\footnotesize CORALIE} spectra. Upper panel: \object{{\footnotesize HD}\,108147}. Lower Panel: \object{{\footnotesize HD}\,168746}. Both objects show low chromospheric activity. Strong activity would result in an emission peak in the center of the absorption line}
\label{fi:ca_line}
\end{figure}

Fig.~\ref{fi:ca_line} shows the \ion{Ca}{ii}\,H absorption line region of the {\footnotesize CORALIE} spectrum at $\lambda 3968.5$\,\AA. The emission flux in the core of the \ion{Ca}{ii}\,H line corrected for the photospheric flux provides us with the chromospheric activity index $S_{\rm Cor}$ \citep{Santos-2000:a} from which we derive the activity indicator $\log{(R^{\prime}_{\rm HK})}=-4.72$. This value is typical for stars with a low chromospheric activity level \citep{Henry-96}. Using the calibration given by \citet{Donahue-93} and quoted in \citet{Henry-96} we compute for this star an age of approximately 2\,Gyrs, while the rotational period resulting from the calibration given by \citet{Noyes-84} is of 8.7\,days. The star is not seen as photometrically variable in the HIPPARCOS data, confirming again the low activity level.
We have estimated the projected rotational velocity of the star to be $v\sin i = 5.3$\,km$s^{-1}$ by means of the {\footnotesize CORALIE} cross-correlation function ({\footnotesize CCF}) \citep{Queloz-98}. The relatively high stellar rotation could cause a small jitter on the radial-velocity data, which are however expected to be in the order of only few m\,s$^{-1}$ \citep{Saar-1997,Santos-2001:a}. The observed and inferred stellar parameters are summarized in Table~\ref{ta:stellar}.

\begin{table}
\caption{
\label{ta:stellar}
Observed and inferred stellar parameters for 
\object{{\footnotesize HD}\,108147} and \object{{\footnotesize
    HD}\,168746}. Photometric and astrometric data have been extracted from the {\footnotesize HIPPARCOS} catalogue \citep{ESA-97} while spectroscopic data are from \citet{Santos-2001:a}}
\begin{tabular}{l@{}lcc}
\hline
  \multicolumn{2}{l}{\bf Parameter} 
& \multicolumn{1}{c}{\bf {\footnotesize HD}\,108147} 
& \multicolumn{1}{c}{\bf {\footnotesize HD}\,168746} \\
\hline
\multicolumn{2}{l}{Spectral Type} & F8/G0 & G5 \\
V     & & 6.99  & 7.95  \\
$B-V$ & & 0.537 & 0.713 \\
$\pi$ &[mas]   & $25.95 \pm 0.69$ & $23.19 \pm 0.96$ \\
Distance &[pc]   & $38.57 \pm 1$ & $ 43.12 \pm 1.8$ \\
$M_V$ & & 4.06 & 4.78\\
$L/L_\odot$ & & 1.93 & 1.10 \\ 
$[Fe/H]$ & & $ 0.2 \pm 0.06 $ & $ -0.06 \pm 0.05 $ \\ 
$M/M_\odot$  & & $1.27 \pm 0.02$ & $0.88 \pm 0.01$ \\ 
$T_{\rm eff}$ & [K] & $6265 \pm 40$ & $5610 \pm 30$ \\
$\log g$ & [cgs] & $4.59 \pm 0.15$ & $4.50 \pm 0.15$ \\
$v\sin i$ & [km\,s$^{-1}$] & $5.3 $ & $1.0 $ \\
$\log(R^{\prime}_{\rm HK})$ & & $-4.72$ &  --  \\
$P_{\rm rot}(R^{\prime}_{\rm HK})$& [days] & 8.7 & -- \\
age($R^{\prime}_{\rm HK}$) & [Gyr] & 2.17 & --/old star \\
\hline
\end{tabular}
\end{table}

\subsection{{\footnotesize CORALIE} orbital solution for 
  \object{{\footnotesize HD}\,108147}}

The precise radial velocity data of \object{{\footnotesize HD}\,108147} have been collected by our group during the period of time from March 1999 to February 2002\footnote{The discovery was announced in May 2000, see www.eso.org/outreach/press-rel/pr-2000/pr-13-00.html}. The result of this campaign is a set of 118 data points having a mean photon-noise error on the individual measurements of $\langle \varepsilon_i \rangle$\,=7.8\,m\,s$^{-1}$. A periodic variation of the radial velocity could be detected on this data, which clearly indicates the presence of a planetary companion. The measured variation cannot be produced by stellar activity: The Geneva photometry data show very low dispersion of 1\,mmag and no CCF-bisector variation \citep{Queloz-2001:b} has been measured at the m\,s$^{-1}$ level. Therefore the planetary explanation seems to be the most likely.
\par
\begin{figure}
\psfig{width=\hsize,file=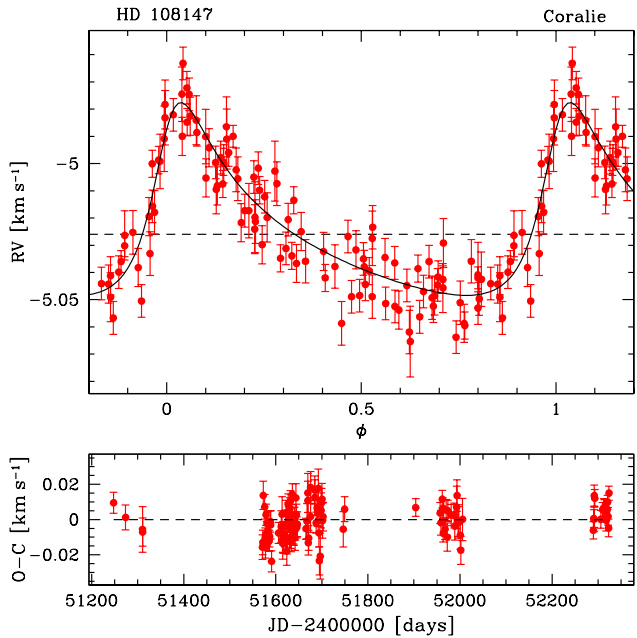}
\caption{
\label{fi:orb108147}
Phase-folded radial-velocity measurements obtained with {\footnotesize CORALIE} for \object{{\footnotesize HD}\,108147}. The error bars represent photon-noise errors only. On the lower panel the residuals of the measured radial velocities to the fitted orbit are plotted as a function of time. They show a tiny indication for the presence of a long-period, second companion of \object{{\footnotesize HD}\,108147}}
\end{figure}

The best-fit Keplerian orbit to the data is shown in Fig.~\ref{fi:orb108147}. It yields a precisely-determined orbital period $P$ of $10.901\pm 0.001$\,days and a large eccentricity \mbox{$e=0.50\pm 0.03$}. The semi-amplitude of the radial-velocity variation is $K=36\pm 1$\,m\,s$^{-1}$. The weighted r.m.s. of the data to the Keplerian fit is 9.2\,m\,s$^{-1}$. The complete set of orbital elements with their uncertainties are given in Table~\ref{ta:orb}.
\par
Using the best-fit orbital parameters and the mass of \object{{\footnotesize HD}\,108147} given above we derive for the companion a {\sl minimum} mass $m_2\,\sin{i}=0.40$\,M$_{\rm Jup}$ (which is about only 1.34 times the mass of Saturn). Because of the many data points and the long observation period the orbit is determined very accurately. This allows us to determine the minimum mass of the companion with accuracy of better than 4\%, provided that we do not consider the major error source, namely the uncertainty on the mass of the primary. From the orbital parameters and the star mass we get also the separation of the companion to its parent star which is $a=0.104$\,AU. The surface equilibrium temperature of the planet at such a distance is estimated to be about 890\,K, following \citet{Guillot-96}.
\par
\object{{\footnotesize HD}\,108147}\,b belongs to the so-called hot Jupiter (or better: hot Saturn!) category of extra-solar planets. The close location to its parent star makes the planet a good candidate for a photometric transit search. The photometric monitoring described in a forthcoming paper (Olsen et al., in prep.) did unfortunately not show any indication for a transit.

\begin{table}
\caption{{\footnotesize CORALIE} best Keplerian orbital solutions derived for 
\object{{\footnotesize HD}\,108147} and \object{{\footnotesize
    HD}\,168746}, as well as inferred planetary parameters}
\label{ta:orb}
\begin{tabular}{l@{}lr@{\hspace{0.05cm}$\pm$\hspace{0.05cm}}lr@{\hspace{0.05cm}$\pm$\hspace{0.05cm}}l}
\hline
  \multicolumn{2}{l}{\bf Parameter} 
& \multicolumn{2}{c}{\bf {\footnotesize HD}\,108147} 
& \multicolumn{2}{c}{\bf {\footnotesize HD}\,168746} \\
\hline
$P$ &$[$days$]$ & 10.901 & 0.001 & 6.403 & 0.001 \\
$T$ &[JD] & 2451591.6 & 0.1 & 2451994.7 & 0.4 \\
$e$ & & 0.498 & 0.025 & 0.081 & 0.029 \\
$\gamma$ &[km\,s$^{-1}$] & 5.026 & 0.001 & -25.562 & 0.001 \\
$\omega$ &[deg] & 318.95  & 3.03 & 16.30 & 20.88 \\
$K$  &[m\,s$^{-1}$] & 36 & 1 & 27 & 1 \\
$N_{\rm meas}$ & & \multicolumn{2}{c}{118} & \multicolumn{2}{c}{154}\\
$\sigma (O$-$C)$\hspace{.25cm}  & [m\,s$^{-1}$] & \multicolumn{2}{c}{9.2} & \multicolumn{2}{c}{9.8} \\
\hline
$a_1\sin i$ & [Mm] & \multicolumn{2}{c}{4.634} & \multicolumn{2}{c}{2.364} \\
$f(m)$ &$\mathrm{[M_{\odot}]}$ 
  & \multicolumn{2}{c}{3.337$\cdot 10^{-9}$} & \multicolumn{2}{c}{1.284$\cdot 10^{-9}$} \\
$m_{1}$ &$\mathrm{[M_{\odot}]}$ 
        & \multicolumn{2}{c}{1.27} & \multicolumn{2}{c}{0.88} \\
$m_{2}\,\sin i$ &$\mathrm{[M_{\rm Jup}]}$ 
        & \multicolumn{2}{c}{0.40} & \multicolumn{2}{c}{0.23} \\
$a$ &[AU] & \multicolumn{2}{c}{0.104} & \multicolumn{2}{c}{0.065} \\
$T_{\rm eq}$ &[K] 
        & \multicolumn{2}{c}{890} & \multicolumn{2}{c}{900} \\
\hline
\end{tabular}
\end{table}

\subsection{A planetary system around \object{{\footnotesize HD}\,108147}\,?}
The weighted r.m.s. of the data to the Keplerian fit is 9.2\,m\,s$^{-1}$ and the reduced $\chi$ is 1.440, indicating an internal error for the single measurements of 6.4\,m\,s$^{-1}$. The residual dispersion is 6.6\,m\,s$^{-1}$ and is only partially composed of instrumental errors, which are in the order of 3\,m\,s$^{-1}$ \citep{Queloz-2001:a}. The remaining 6\,m\,s$^{-1}$ could be due to stellar jitter, since the star is fairly rapidly rotating. Closer analysis of the residuals (Fig.~\ref{fi:orb108147}, bottom) might indicate however the presence of a possible second companion with an orbital period $P$ of $587 \pm 34$\,days and a minimum mass of $m_2\,\sin{i}=0.42$\,M$_{\rm Jup}$. A best-fit solution considering a second companion reduces the r.m.s to the Keplerian to 8.0\,m\,s$^{-1}$ while the reduced $\chi$ passes from 1.440 to 1.290. In order to confirm or reject this possibility we will continue to collect additional high-precision data during the next observational seasons. We plan to monitor the radial velocity of this object using the {\footnotesize CORALIE} spectrograph, but also taking advantage of the superior performances of the {\footnotesize HARPS}\footnote{HARPS will be installed on the ESO 3.6-m telescope and become operational at the beginning of 2003} spectrograph \citep{Pepe-2000} which is expected to reach a radial-velocity precision of 1\,m\,s$^{-1}$.

\subsection{Results of the improved cross correlation}
As mentioned previously, the radial velocity data have been computed by using the weighted cross correlation and the new numerical mask. For comparison we have nevertheless computed the radial velocity data also using the standard on-line data reduction. The results are summarized in Table~\ref{ta:comp}. 
\par
In the error budget of our best single-orbit solution about 6.6\,m\,s$^{-1}$ arise from sources other that photon noise. The r.m.s obtained with the standard algorithm was 11.9\,m\,s$^{-1}$ and with the weighted cross correlation 10.1\,m\,s$^{-1}$. Considered this, the weighted cross correlation reduced the radial velocity dispersion arising from photon noise by a factor 1.31, confirming the simulation results. On the other hand, the r.m.s of 11.9\,m\,s$^{-1}$ obtained with the original mask passes to 10.7\,m\,s$^{-1}$ with the new clean mask. Thus, we estimate that the telluric lines add a dispersion on the stellar radial velocity of roughly 5\,m\,s$^{-1}$.

\begin{table}
\caption{Comparison of the different data reductions on \object{{\footnotesize HD}\,108147}. The r.m.s of the data to the Keplerian fit is reduced from 11.9 to 9.2\,m\,s$^{-1}$ if the weighted cross correlation and the clean numerical mask are used}
\label{ta:comp}
\begin{tabular}{l c }
\hline
{\bf Cross correlation}	&	{\bf Obtained r.m.s [m\,s$^{-1}$]}\\
\hline
Standard	&	11.9\\
Clean mask only	&	10.7\\
Weighted only	&	10.1\\
Weighted \& clean mask	& 9.2\\
\hline
\end{tabular}
\end{table}

\subsection{The eccentricity of the orbit around \object{{\footnotesize HD}\,108147}}
Considered the short orbital period, the eccentricity of the orbit around \object{{\footnotesize HD}\,108147} is surprisingly high. According to the commonly believed migration scenario for the formation of hot Jupiters, the giant planets like \object{{\footnotesize HD}\,108147} cannot be formed {\it in situ} at a distance of only 0.1\,AU from the parent star. On the other hand the orbit of planets having suffered a strong migration are expected to be circularized by the tidal interaction with the accretion disk \citep{Goldreich-1980}. Another mechanism must therefore be at the origin of its high eccentricity. One suitable explanation is provided by the gravitational interaction between multiple giant planets in the early stages of the system formation \citep{Weidenschilling-96,Rasio-1996:a,Lin-97}. As a result one planet could have been projected on a short-period orbit with a high eccentricity, leaving different possibilities concerning the destiny of its scattering partner. Eccentric orbits can however also be produced by additional companions on longer-period orbits, which perturb the short-period orbit of the inner companion by tidal forces \citep{Mazeh-97:a}. The companion can be either planetary or stellar and would be seen in many cases only as a slow drift of the gamma-velocity $\gamma$. Finally, additional scenarios to explain high eccentricities were proposed recently: For example the migration of a single Jupiter-mass planet through a planetesimal disk \citep{Levison-98}, or, the simultaneous formation and migration of \emph{two} or more planets through the accretion disk, e.g locked in a resonant system \citep{Murray-2002}. The numerical simulations in the latter paper show that a large variety of possible results can be obtained, depending on the initial planetary mass(es), disk properties and time scales.
\par
In summary, various mechanisms can explain short-period eccentric orbits. In almost all of them interaction with planetary or stellar companions plays a fundamental role, and the large eccentricity of the orbit might be an indication for their presence. Therefore it is very likely to find a second, long-period companion around \object{{\footnotesize HD}\,108147}. As mentioned above the long-term follow up of this object might confirm or reject this possibility.

\section{A Sub-Saturn mass planet around \object{{\footnotesize HD}\,168746}}

\subsection{Stellar characteristics of \object{{\footnotesize HD}\,168746}}

{\footnotesize HD}\,168746 ({\footnotesize HIP}\,90004) is a G5 dwarf of magnitude $V$\,=\,7.95 located at the boundary between the Scutum and the Serpens Cauda constellations. The {\footnotesize HIPPARCOS} catalogue lists a color index $B-V$\,=\,0.713. The catalogue indicates also an astrometric parallax of $\pi=23.19$\,mas \citep{ESA-97} corresponding to a distance of about 38.57\,pc between the star and the Sun. The resulting absolute magnitude is $M_V$\,=\,4.78.
\par
The spectroscopic analysis by \citet{Santos-2001:a} indicates an effective temperature $T_{\rm eff}=5610$\,K, $\log g=4.50$, and [Fe/H]\,=\,$-0.06$. The combination of the precise spectroscopic parameters with the evolution models provide a stellar mass of $M=0.88\,M_{\odot}$. The emission flux in the core of the \ion{Ca}{ii}\,H shows no evidence for chromospheric activity (see Fig.~\ref{fi:ca_line}). Since the star is too faint it has not been possible to compute a reliable value for $R^{\prime}_{\rm HK}$ from the {\footnotesize CORALIE} spectra. Therefore it was not possible to derive neither the age nor $P_{\rm rot}$. Evolutionary tracks suggest for this star an age of at least several Gyrs.

\subsection{{\footnotesize CORALIE} orbital solution for 
\object{{\footnotesize HD}\,168746}}

During the period spanning from May 1999 to September 2001\footnote{The discovery was announced in May 2000, see www.eso.org/outreach/press-rel/pr-2000/pr-13-00.html} we have recorded 154 precise radial-velocity data points of \object{{\footnotesize HD}\,168746}. Because of the magnitude of the star the mean photon-noise error on the individual measurements is relatively high, namely of $\langle \varepsilon_i \rangle$\,=9.7\,m\,s$^{-1}$. Due to the large amount of data we could nevertheless identify easily the presence of a companion with a very low minimum mass. Because of the photometric stability of \object{{\footnotesize HD}\,168746} and the absence of line-bisector variations we can exclude that the measured radial-velocity variation is produced by stellar activity.

\begin{figure}
\psfig{width=\hsize,file=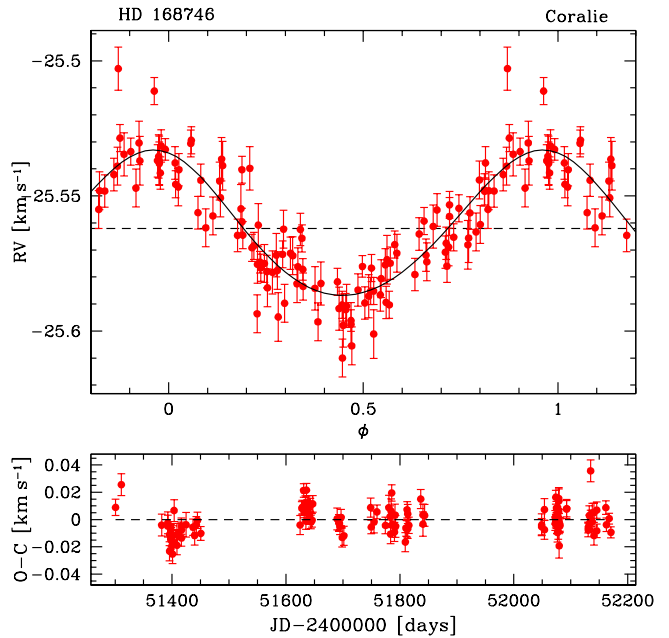}
\caption{Phase-folded radial-velocity measurements obtained with {\footnotesize CORALIE} for \object{{\footnotesize HD}\,168746}. The error bars represent photon-noise errors only. The lower panel shows the residuals of the measured radial velocities to the fitted orbit as a function of time}
\label{fi:orb168746}

\end{figure}

Fig.~\ref{fi:orb168746} shows the best fitted Keplerian orbit for the companion of \object{{\footnotesize HD}\,168746}. We deduce an orbital period $P$ of $6.403\pm 0.001$\,days and a rather small eccentricity of $e=0.08\pm 0.03$. The semi-amplitude of the radial-velocity variation is only $K=27\pm 1$\,m\,s$^{-1}$. The period and the semi-amplitude are well constrained thanks to the large amount of data points covering many periods. The complete set of orbital elements with their uncertainties are given in Table~\ref{ta:orb}.
\par
The best-fit orbital parameters and the mass of \object{{\footnotesize HD}\,168746} yield for its companion a {\sl minimum} mass of $m_2\,\sin{i}=0.23$\,M$_{\rm Jup}$, which is only 0.77 times the mass of Saturn. The accuracy obtained on this mass estimation is better than 3\%, if we again do not consider the uncertainty on the mass of the primary. The separation of the companion to its parent star is $a=0.065$\,AU. The surface equilibrium temperature of the planet at such a distance is about 900\,K. Similarly to \object{{\footnotesize HD}\,108147}, a dedicated search for a possible photometric transit of the planets in front of \object{{\footnotesize HD}\,168746} remained unsuccessful (Olsen et. al, in prep.).
\par
The radial velocity data were computed, as for \object{{\footnotesize HD}\,108147}, using the weighted correlation and the cleaned numerical mask. We confirm a reduction of the measured dispersion: The weighted r.m.s. of the data to the Keplerian fit decreased from the original 11.2\,m\,s$^{-1}$ to the final 9.8\,m\,s$^{-1}$, with a reduced $\chi$ of 1.5.
\par
The {\footnotesize CORALIE} individual radial-velocity measurements presented in this paper are available in electronic form at {\footnotesize CDS} in Strasbourg.

\section{Concluding remarks}
The companions to \object{{\footnotesize HD}\,108147} and \object{{\footnotesize HD}\,168746} possess minimum masses of 1.34 and 0.77\,M$_{\rm Sat}$, respectively. The latter is one of the lightest extra-solar planet candidates found to date. Both objects have been detected by means of the {\footnotesize CORALIE} spectrograph. {\footnotesize CORALIE} and its twin instrument {\footnotesize ELODIE} are particularly efficient for the detection of light planets: Seven of the ten lightest candidates and two of the four sub-saturnian companions have been discovered using these instruments\footnote{obswww.unige.ch/$\sim$naef/who\_discovered\_that\_planet.html}, among which \object{{\footnotesize HD}\,83443}\,c, the lightest ever discovered planetary companion with only 0.53\,M$_{\rm Sat}$.
\par
The suitability of {\footnotesize CORALIE} for the detection of low-mass planetary companions arises from the intrinsic instrument precision which is better than 3\,m\,s$^{-1}$ over long time scales. Despite the modest size of the telescope the overall detection efficiency of {\footnotesize CORALIE} is high, due certainly to the telescope-time availability and, in particular, to the simultaneous ThAr reference. The ThAr technique allows us to cover the full visible wavelength range without any transmission losses, ensuring a gain in flux of at least a factor of 6 compared to the use of an iodine absorption cell. The photon noise is thus reduced by a factor of about 2-2.5 compared to an exposure of same duration using the iodine cell \citep{Bouchy-2001b}. Nevertheless, the measurements remain in many cases photon-noise limited. The new weighted cross-correlation algorithm presented in this paper allowed us to reduce by about 1.25 the photon-noise contribution to the radial-velocity data. Although some improvement of {\footnotesize CORALIE} and its data reduction software is still going on, our major future step will be to complete the realization of {\footnotesize HARPS} \citep{Pepe-2000}, a new stable and efficient echelle spectrograph dedicated to the search for extra-solar planets with the ESO 3.6-m telescope at La Silla. The aimed precision of 1\,m\,s$^{-1}$ should considerably increase the detection rate of sub-saturnian planets.

\begin{acknowledgements}
We thank Dr. Paul Bartholdi for his precious help in implementing the "weighted correlation" method. We are grateful to the staff from the Geneva and Haute-Provence Observatories who have built and maintain the new 1.2-m Euler Swiss telescope and the CORALIE echelle spectrograph at La\,Silla. We thank the Geneva University and the Swiss NSF (FNRS) for their continuous support for this project.  Support to N.S.  from Funda\c{c}\~ao para a Ci\^encia e Tecnologia (Portugal) is gratefully acknowledged.
\end{acknowledgements}

\bibliography{H3477}
\bibliographystyle{aa}

\end{document}